\documentclass{article}
\pdfoutput=1

\usepackage{arxiv}
\usepackage[utf8]{inputenc} 
\usepackage[T1]{fontenc}    
\usepackage{url}            
\usepackage{booktabs}       
\usepackage{amsfonts}       
\usepackage{nicefrac}       
\usepackage{microtype}      
\usepackage{lipsum}
\usepackage{graphicx}
\usepackage{xcolor}
\usepackage{enumitem}
\usepackage{amsmath}
\usepackage{booktabs,graphicx,makecell,siunitx,eqparbox}

\title{Data-driven discovery of dimensionless numbers and scaling laws from experimental measurements}

\author{
 Xiaoyu Xie \\
    Department of Mechanical Engineering\\
    Northwestern University\\
    Evanston, IL 60208 \\
    \texttt{xiaoyuxie2020@u.northwestern.edu} \\
   \And
 Wing Kam Liu\thanks{Corresponding authors: w-liu@northwestern.edu (W. Liu), zhengtao.gan@northwestern.edu (Z. Gan)} \\
	Department of Mechanical Engineering\\
	Northwestern University\\
	Evanston, IL 60208 \\
	\texttt{w-liu@northwestern.edu} \\  
    \And   
 Zhengtao Gan \footnotemark[1] \\
    Department of Mechanical Engineering\\
    Northwestern University\\
    Evanston, IL 60208 \\
    \texttt{zhengtao.gan@northwestern.edu} \\
}

\begin{document}
\maketitle
\begin{abstract}
Dimensionless numbers and scaling laws provide elegant insights into the characteristic properties of physical systems. Classical dimensional analysis and similitude theory fail to identify a set of unique dimensionless numbers for a highly-multivariable system with incomplete governing equations. In this study, we embed the principle of dimensional invariance into a two-level machine learning scheme to automatically discover dominant and unique dimensionless numbers and scaling laws from data. The proposed methodology, called dimensionless learning, can reduce high-dimensional parametric spaces into descriptions involving just a few physically-interpretable dimensionless parameters, which significantly simplifies the process design and optimization of the system. We demonstrate the algorithm by solving several challenging engineering problems with noisy experimental measurements (not synthetic data) collected from the literature. The examples include turbulent Rayleigh-Bénard convection, vapor depression dynamics in laser melting of metals, and porosity formation in 3D printing. We also show that the proposed approach can identify dimensionally-homogeneous differential equations with minimal parameters by leveraging sparsity-promoting techniques.

\end{abstract}

\keywords{Dimensional analysis \and Buckingham's Pi theorem \and Physics-informed machine learning \and Fluid mechanics \and Sparse identification \and Additive manufacturing}

\section{Introduction}


All physical laws can be expressed as dimensionless relationships with fewer dimensionless numbers and a more compact form \cite{barenblatt2003scaling}. A dimensionless number is a power-law monomial of some physical quantities \cite{tan2011dimensional}. There is no physical dimension (such as mass, length, or energy) assigned to a dimensionless number, which provides a property of scale invariance, i.e., the dimensionless numbers are invariant when length scale, time scale, or energy scale of the system varies. More than 1200 dimensionless numbers have been discovered in an extremely wide range of fields, including physics and physical chemistry, fluid and solid mechanics, thermodynamics, electromagnetism, geophysics and ecology, and engineering \cite{kunes2012dimensionless}. There are several significant advantages to describe a physical process or system using dimensionless numbers. Using dimensionless numbers can considerably simplify a problem by reducing the number of variables that describe the physical process, thereby reducing the amount of experiments (or simulations) required to understand and design the physical system. For example, Reynolds number ($\mathrm{Re}$) is a well-known dimensionless number in fluid mechanics, named after Osborne Reynolds who investigated the fluid flow through pipes in 1883 \cite{reynolds1883xxix}. The Reynolds number is defined as a power law of four physical quantities, i.e., the fluid density, the average fluid velocity, the diameter of the pipe, and the dynamic fluid viscosity. The flow characteristics (laminar or turbulent) in a pipe is best determined by the $\mathrm{Re}$, rather than the four individual dimensional quantities. Moreover, the scale-invariance property of the dimensionless numbers plays a critical role in the similitude theory \cite{kline2012similitude}. Many small-scale experiments have been designed to understand and predict the behaviors of the full-scale applications in the engineering of aerospace \cite{ghosh1980large}, nuclear \cite{nahavandi1979scaling}, and marine \cite{vassalos1998physical}, where the full-scale applications are typically extremely expensive even dangerous. All the dimensionless numbers should be identical between the small-scale and full-scale experiments which yields the perfect similarities of geometry, dynamic, and kinematic between the two scales. Furthermore, dimensionless numbers are ratios of two forces, energies, or mechanisms. Thus, they are physically interpretable and can provide fundamental insights into the behavior of complex systems. For example, The Péclet number ($\mathrm{Pe}$) represents the ratio of the convection rate of a physical quantity by the flow to the gradient-driven diffusion rate, which enables the order-of-magnitudes analysis for transport phenomena of a process.

Despite the scientific significance and the wide use of dimensionless numbers, it is still challenging to discover new dimensionless numbers and their relationships (i.e., scaling laws) from experiments, especially for a complex physical system without complete governing equations. A traditional solution is dimensional analysis \cite{tan2011dimensional} based on Buckingham $\pi$ theorem \cite{buckingham1914physically}, which provides a systematic approach to examine the units of a physical system and form a set of dimensionless numbers that satisfy the principle of dimensional invariance \cite{osborne1978dimensional}. However, the dimensional analysis has several well-known limitations. First, the derived dimensionless numbers are not unique. The Buckingham $\pi$ theorem \cite{buckingham1914physically}, from the viewpoint of mathematics, provides a linear subspace of exponents that produces dimensionless numbers. Any basis for the subspace is equally valid. Thus, it fails to identify the dimensionless numbers, given a particular choice of basis, that are dominant for the physical system. Second, the mathematical relation between dimensionless numbers (i.e., scaling law) remains unknown by using dimensional analysis alone. A common approach to establish the scaling law is to leverage the results of the dimensional analysis with experimental measurements of the physical system. The experimental measurements are transformed into the dimensionless numbers obtained by dimensional analysis and fitted a high-dimensional response surface to represent the scale-invariant relationship. However, since the dimensional analysis does not provide unique dimensionless numbers, this procedure is very time-consuming and highly relies on the experience of domain experts to select a set of appropriate dimensionless numbers by a lot of trial-and-error. 

Theses limitations could be overcame by integrating the dimensional analysis with advanced data science and artificial intelligence (AI). Mendez and Ordonez proposed an algorithm called SLAW (i.e., Scaling LAWs) to identify the form of a power law from experimental data (or simulation data) \cite{mendez2005scaling}. The proposed SLAW combines dimensional analysis with multivariate linear regressions. This approach has been applied to some engineering areas, such as ceramic-to-metal joining \cite{mendez2005scaling} and plasma confinement in Tokamaks \cite{murari2015new}. However, for the sake of simplification, the proposed algorithm assumes the relationship between the dimensionless numbers obeys a power law, which is invalid in many applications. Constantine, Rosario, and Iaccarino proposed a rigours mathematical framework to estimate unique and relevant dimensionless groups \cite{constantine2017data,constantine2016many}. Active subspace methods are connected to dimensional analysis, which reveals that all physical laws are ridge functions \cite{constantine2016many}. Their method is applicable to idealized physical system meaning that the experiments can be conducted for arbitrary values of the independent input variables (or dependent input variables with a known probability density function), and noises or errors in the input and output are negligible. 

In this study, we propose a mechanistic data-driven approach, called dimensionless learning, in which the principle of dimensional invariance (i.e., physical laws cannot depend on an arbitrary choice of basic units of measurements \cite{barenblatt2003scaling}) is embedded into a two-level machine learning scheme to automatically discover dominant dimensionless numbers and scaling laws from noisy experimental measurements of complex physical systems. We add a physical constraint of the dimensional invariance into the learning algorithm, which steers the learning toward scale-invariant and physically-interpretable low-dimensional patterns of complex high-dimensional systems. We demonstrate the proposed approach by solve multiple challenging problems in science and engineering. The examples include turbulent Rayleigh-Benard convection, vapor depression dynamics, and porosity formation in 3D printing. The experimental datasets are collected from the literature. Furthermore, we demonstrate the potential of the proposed approach by combining it with the sparsity-promoting techniques (such as SINDy \cite{brunton2016discovering}) to identify dimensionless differential equations from data, which involves natural coordinates of spatial and temporal variables.    

\section{Results}

\subsection{Turbulent Rayleigh-Bénard convection}

\begin{figure}
  \centering
  \includegraphics[width=1\linewidth]{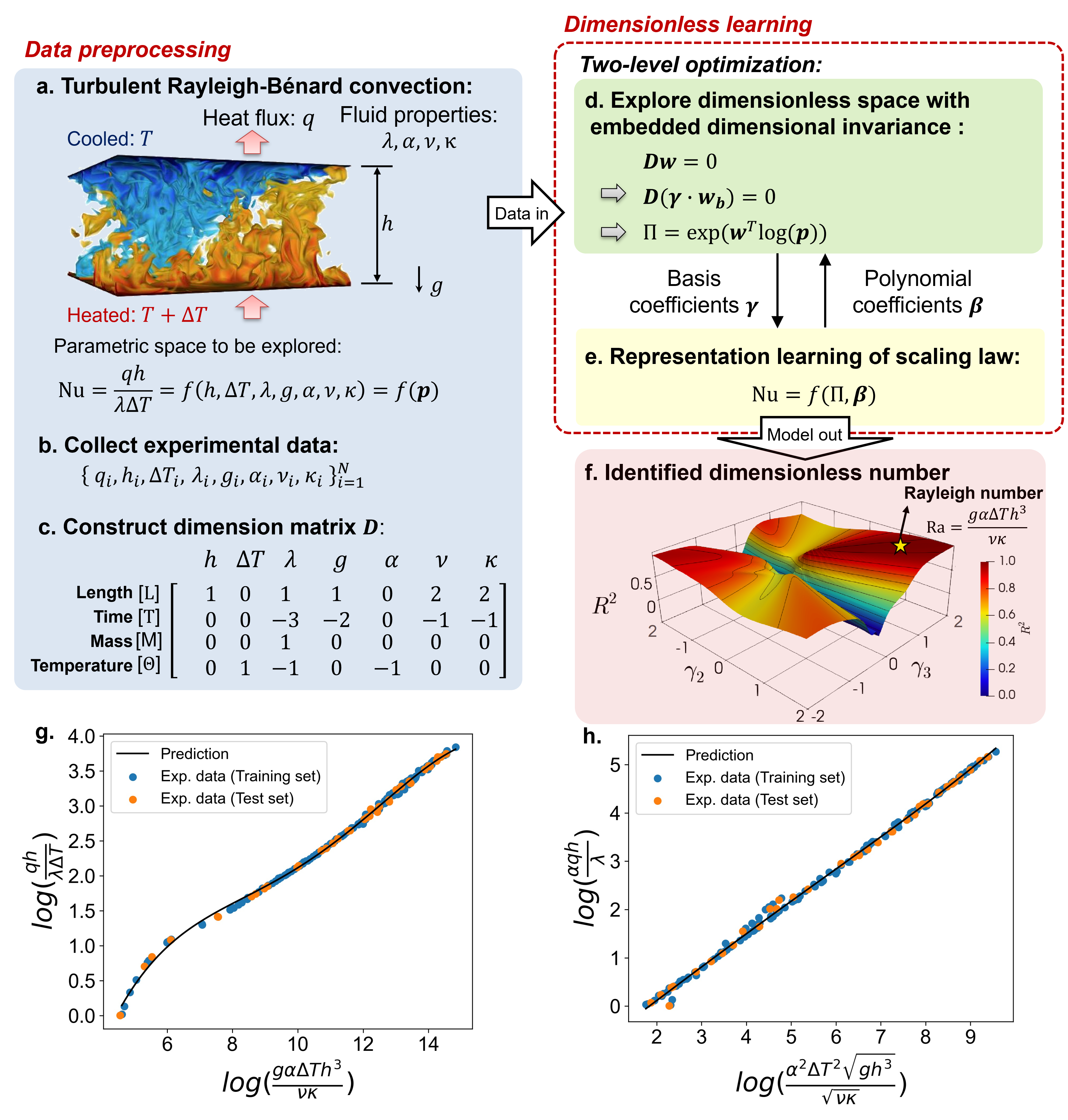}
  \caption{\textbf{The proposed dimensionless learning demonstrated on the turbulent Rayleigh-Benard convection.} \textbf{(a)} A schematic of the Rayleigh-Benard convection \cite{RByoutube} with related physical quantities. \textbf{(b)} Collected experimental measurements. \textbf{(c)} Constructed dimension matrix $\boldsymbol{D}$ of the input variables. \textbf{(d)} First-level of the two-level optimization scheme for training the coefficients $\boldsymbol{\gamma}$ with respect to the computed basis vectors. \textbf{(e)} Second-level of the two-level optimization scheme for optimizing the unknown coefficients $\boldsymbol{\beta}$ in the representation learning. \textbf{(f)} Explored dimensionless space with a measure of $R^2$. The location with the maximum $R^2$ marked as a yellow star is related to the classical Rayleigh number. \textbf{(g)} Identified one-dimensional scaling law between Nu and Ra. \textbf{(h)} Discovered linear scaling law between two identified dimensionless numbers.}
  \label{fig:rayleigh_results}
\end{figure}

In this section, we demonstrate the proposed dimensionless learning using the example of a classical fluid mechanics problem: turbulent Rayleigh-Bénard convection. The goal is to rediscover the Rayleigh number (Ra) directly from experimental measurements. The Ra is named after Lord Rayleigh who investigated a non-isothermal buoyancy-driven flow in 1916 \cite{rayleigh1916lix}, which is named Rayleigh-Bénard convection now. The turbulent Rayleigh-Bénard convection is a paradigm system to study turbulent thermal flow, occurring in a planar horizontal layer of fluid in a container heated from below. The internal fluid could develop complex turbulent dynamics due to the effects of buoyancy, fluid viscosity, and gravity (Fig. \ref{fig:rayleigh_results}a).

Experimentally, the heat flux through the container, $q$, can be measured, which depends on the height of the container $h$, the temperature difference between the top and bottom surfaces $\Delta T$, the gravitational acceleration $g$, and the properties of the fluid including the thermal conductivity $\lambda$, the thermal expansion coefficient $\alpha$, the viscosity $\nu$, and the thermal diffusivity $\kappa$. To obtain a causal relationship, we need to specify the dependent (i.e., output) and independent (i.e., input) variables from the physical quantities describing the system. To simplify the demonstrate, we assume the form of the output variable as the Nusselt number $\mathrm{Nu}=\frac{q h}{\lambda \Delta T}$ (a more general case using $q$ as the output will be presented later), and a list of physical quantities $\boldsymbol{p}$ as input variables. The causal relationship to be determined can be represented as 
\begin{equation}
	\mathrm{Nu}=\frac{q h}{\lambda \Delta T} = f(h, \Delta T, \lambda, g, \alpha, \nu, \kappa) = f(\boldsymbol{p})
\end{equation}

It is a high-dimensional parametric space. To explore it, we collect an experimental dataset of the turbulent Rayleigh-Bénard convection from two different articles \cite{niemela2006turbulent, chavanne2001turbulent}, including 182 experiments with various input variables and corresponding output measurements (Fig. \ref{fig:rayleigh_results}b). Tons of machine learning models can fit the data. However, a lot of those are black box models, such as neural networks, without good interpretability and physical insights. Alternatively, we aim to identify a low-dimensional scale-invariant scaling law that best represent the dataset. In the scaling law, the products of powers of the input variables $\boldsymbol{p}$ form a dimensionless number $\mathrm{\Pi}$. Thus, the causal relationship can be rewritten as
\begin{equation}
	\mathrm{Nu}=f_1(\mathrm{\Pi})
\end{equation}
\begin{equation}
	\mathrm{\Pi}=h^{w_1}\Delta T^{w_2}\lambda^{w_3}g^{w_4}\alpha^{w_5}\nu^{w_6}\kappa^{w_7}
	\label{equ:pi}
\end{equation}
where $\boldsymbol{w}=[w_1,...,w_7]^T$ are the powers that produce the dimensionless number and are to be determined. In this example, we assume that there is only one input dimensionless number governing the process. An algorithm for determining the number of the dimensionless numbers required from data is provided in Supplementary Information Section 1.3.

To embed the physical constraint of dimensional invariance, we conduct the dimensional analysis, i.e., the powers $\boldsymbol{w}=[w_1,...,w_7]^T$ need to satisfy the a linear system of equations
\begin{equation}
	\boldsymbol{Dw} =0
	\label{equ:linear system}
\end{equation}

where $\boldsymbol{D}$ is the dimension matrix of the input variables (Fig. \ref{fig:rayleigh_results}c). Each column of the dimension matrix is the dimension vector of the corresponding variable. The dimension vector represents the exponents of the physical quantity with respect to the fundamental dimensions. It is worthy noting that there are only seven fundamental dimensions in nature: mass [$\mathrm{M}$], length [$\mathrm{L}$], time [$\mathrm{T}$], temperature [$\mathrm{\Theta}$], electric current [$\mathrm{I}$], luminous intensity [$\mathrm{J}$], and amount of substance [$\mathrm{N}$] \cite{gHobel2006international}. All the other dimensions are power-law monomials of the fundamental dimensions \cite{barenblatt2003scaling}. In this example, we use four fundamental dimensions, i.e., [$\mathrm{M}$], [$\mathrm{L}$], [$\mathrm{T}$], and [$\mathrm{\Theta}$] (Fig. \ref{fig:rayleigh_results}c). The dimension matrix includes the physical dimensions of the input variables. The linear system of equations $\boldsymbol{Dw} =0$ guarantees the power-law monomial of the input variables (Eqn. \ref{equ:pi}) is dimensionless \cite{calvetti2015dimensional}. Since the linear system is under-determined (i.e., the number of unknown variables is more than the number of equations), there are infinitely many solutions, indicating that there are infinitely many forms of dimensionless numbers obtained from the dimensional analysis. Furthermore, we can represent the solutions of the linear system (Eqn.\ref{equ:linear system}) as linear combinations of three basis vectors $\boldsymbol{w_{b1}}$, $\boldsymbol{w_{b2}}$, and $\boldsymbol{w_{b3}}$ 
\begin{equation}
	\boldsymbol{w} = \gamma_1 \boldsymbol{w_{b1}} + \gamma_2 \boldsymbol{w_{b2}} + \gamma_3 \boldsymbol{w_{b3}}
	\label{equ:basisvectors}
\end{equation}

where $\boldsymbol{\gamma}=[\gamma_1, \gamma_2, \gamma_2]^T$ are the coefficients with respect to the three basis vectors in this case. The number of basis vectors is equal to the number of the input variables (seven in this case) minus the rank of the dimension matrix (four in this case). This formula aligns with the Buckingham $\pi$ theorem \cite{buckingham1914physically}. Since the basis vectors can be computed based on Eqn. \ref{equ:linear system} (an algorithm for computing basis vectors is provided in Supplementary Information Section 1.5), the basis vectors' coefficients (or just called basis coefficients) are the unknowns to be determined. A set of computed basis vectors for this case are
\begin{equation}
	\boldsymbol{w_{b1}} = [0,0,0,0,0,1,-1]^T
	\label{equ:basisvectors1}
\end{equation}
\begin{equation}
	\boldsymbol{w_{b2}} = [0,1,0,0,1,0,0]^T
	\label{equ:basisvectors2}
\end{equation}
\begin{equation}
	\boldsymbol{w_{b3}} = [3,0,0,1,0,-2,0]^T
	\label{equ:basisvectors3}
\end{equation}

Once the basis coefficients $\gamma_1$, $\gamma_2$, and $\gamma_3$ are obtained, the form of the dimensionless number $\mathrm{\Pi}$ could be determined by Eqns. \ref{equ:pi} and \ref{equ:basisvectors} (Fig. \ref{fig:rayleigh_results}d).

To determine the values of basis coefficients using the collected dataset, a model representing the scaling relation between the input and output dimensionless numbers is required, which introduces another set of unknown parameters $\boldsymbol{\beta}$ (i.e., the representation learning shown in Fig. \ref{fig:rayleigh_results}e). We use a 5-ordered polynomial model in this case (more advanced models such as tree-based models and deep neural networks are optional depending on the complexity of the problem to be solved, see Section \ref{porosity_sec} and Supplementary Information Section Section 3 for more demonstrations). The polynomial model can be expressed as
\begin{equation}
    \mathrm{Nu} = \beta_0 + \beta_1 \Pi + \beta_2 \Pi^2 + ... + \beta_5 \Pi^5
\end{equation}

where $\boldsymbol{\beta}=[\beta_0, \beta_1,..., \beta_5]^T$ are polynomial coefficients that represents the scaling relation. 

We design an iterative two-level optimization scheme to determine the two sets of unknown parameters in the regression problem, i.e., the basis coefficients $\boldsymbol{\gamma}$ and polynomial coefficients $\boldsymbol{\beta}$. The optimization scheme includes multiple interactive steps. At each step, we adjust the first-level basis coefficients $\boldsymbol{\gamma}$ while fix the second-level polynomial coefficients $\boldsymbol{\beta}$, and then optimize the second-level polynomial coefficients $\boldsymbol{\beta}$ while fix the first-level basis coefficients $\boldsymbol{\gamma}$. This process is repeated until the result is converged, i.e., the values of $\boldsymbol{\gamma}$ and $\boldsymbol{\beta}$ are unchanged. There are several advantages of the proposed two-level approach over a single-level approach combining the two sets of unknowns together during the optimization. We can use different optimization methods and parameters (such as the learning rate) for these two-level models to significantly improve the efficiency of the optimization. More importantly, we can utilize the physical insights to inform the learning process. The first-level basis coefficients $\boldsymbol{\gamma}$ have clear physical meaning, which is related to the powers that produce the dimensionless number. Thus, those values have to be rational numbers to maintain the dimensional invariance. Moreover, the range of them are typically not large. Note that absolute values of coefficients in most of the dimensionless numbers and scaling laws are less than four \cite{barenblatt2003scaling}. To leverage those physical insights or constraints, we design several methods for optimizing the first-level basis coefficients, including a simple grid search (used in this section) and a pattern search that is much more efficient (Supplementary Information Section 3.2). For the second-level coefficients, we conduct multiple standard representation learning methods, including polynomial regression used in this section, tree-based extreme gradient boosting (XGBoost \cite{chen2016xgboost}) used in Section \ref{porosity_sec}, and general gradient descent method (Supplementary Information Section 3.1). Details of the two-level optimization framework is provided in Supplementary Information Section 3.

We illustrate the first-level grid search for $\gamma_2$ and $\gamma_3$ ranging from -2 to 2 with a interval of 1/100 (Fig. \ref{fig:rayleigh_results}f). We fix $\gamma_1$ as one to avoid too large or too small powers of the dimensionless number $\mathrm{\Pi}$. For each $\boldsymbol{\gamma}$ in the dimensionless space, the polynomial coefficients $\boldsymbol{\beta}$ are trained based on the collected data. The data is separated into 80\% training set and 20\% test set. The coefficient of determination ($R^2$) of test set is shown in Fig. \ref{fig:rayleigh_results}f as a measure of learning performance. We can identify an unique point with the maximum $R^2$ (0.999) from Fig. \ref{fig:rayleigh_results}f (marked as a yellow star), where $\gamma_1$=$\gamma_2$=$\gamma_3$ = 1. Using these optimized basis coefficients, the expression of the dominant dimensionless number can be identified as 
\begin{equation}
    \mathrm{\Pi}=\frac{g\alpha \Delta T h^3}{\nu \kappa}
\end{equation}

This form is identical as the classical Rayleigh number, indicating that the proposed dimensionless learning can rediscover the well-known dimensionless number directly from data. Moreover, we demonstrate that for the given parameter list the Rayleigh number is the unique dimensionless number to best fit the dataset because there is only one global maximum of $R^2$ within the dimensionless space (Fig. \ref{fig:rayleigh_results}f). The log-log scaling relation between Ra and Nu is a simple one-dimensional pattern where all the datapoints collapse onto a single curve (Fig. \ref{fig:rayleigh_results}g). 

The proposed dimensionless learning can deal with dimensional output variable as well. A combination of the input variables with the same dimension as the output variable can be searched to non-dimensionalize the output variable (the detailed algorithm for output non-dimensionalization is provided in Supplementary Information Section 1.3). Using the heat flux $q$ as the output variable (instead of Nu used in the previous example), the dimensionless space is expanded and thus more dominant dimensionless numbers and scaling laws can be found. We discover a new set of dimensionless numbers to best represent ($R^2$ = 0.999) the collected experimental measurements. More interestingly, the identified log-log scaling relation between the dimensionless numbers is almost linear (Fig. \ref{fig:rayleigh_results}h). This finding could lead to new physical insights into the complex turbulent Rayleigh-Bénard dynamics.

\subsection{Vapor depression dynamics in laser-metal interaction}
\label{keyhole_sec}

Another challenging problem in the application of dimensionless learning is the laser-metal interaction dynamics. People have been curious about the physical responses of a metallic material due to a high-power laser irradiation since 1964 when Patel invited an electric discharge $\mathrm{CO_2}$ laser \cite{patel1964continuous} that was dramatically scaled up in power shortly after. During the laser-metal interaction, a vapor-filled depression (termed a keyhole) frequently forms on a puddle of liquid metal melted by the laser. The keyhole is caused by vaporization-induced recoil pressure, and its dynamics is inherently difficult to understand because of its complex dependence upon many physical mechanisms but important to be able to quantify because it is highly related to energy absorption and defect formation in many industrial and military applications, such as laser-based materials processing and manufacturing \cite{zhao2019bulk}, high-energy laser weapons \cite{cook2012high}, and aerospace laser-propulsion engine \cite{pirri1972laser}.

\begin{figure}
  \centering
  \includegraphics[width=1\linewidth]{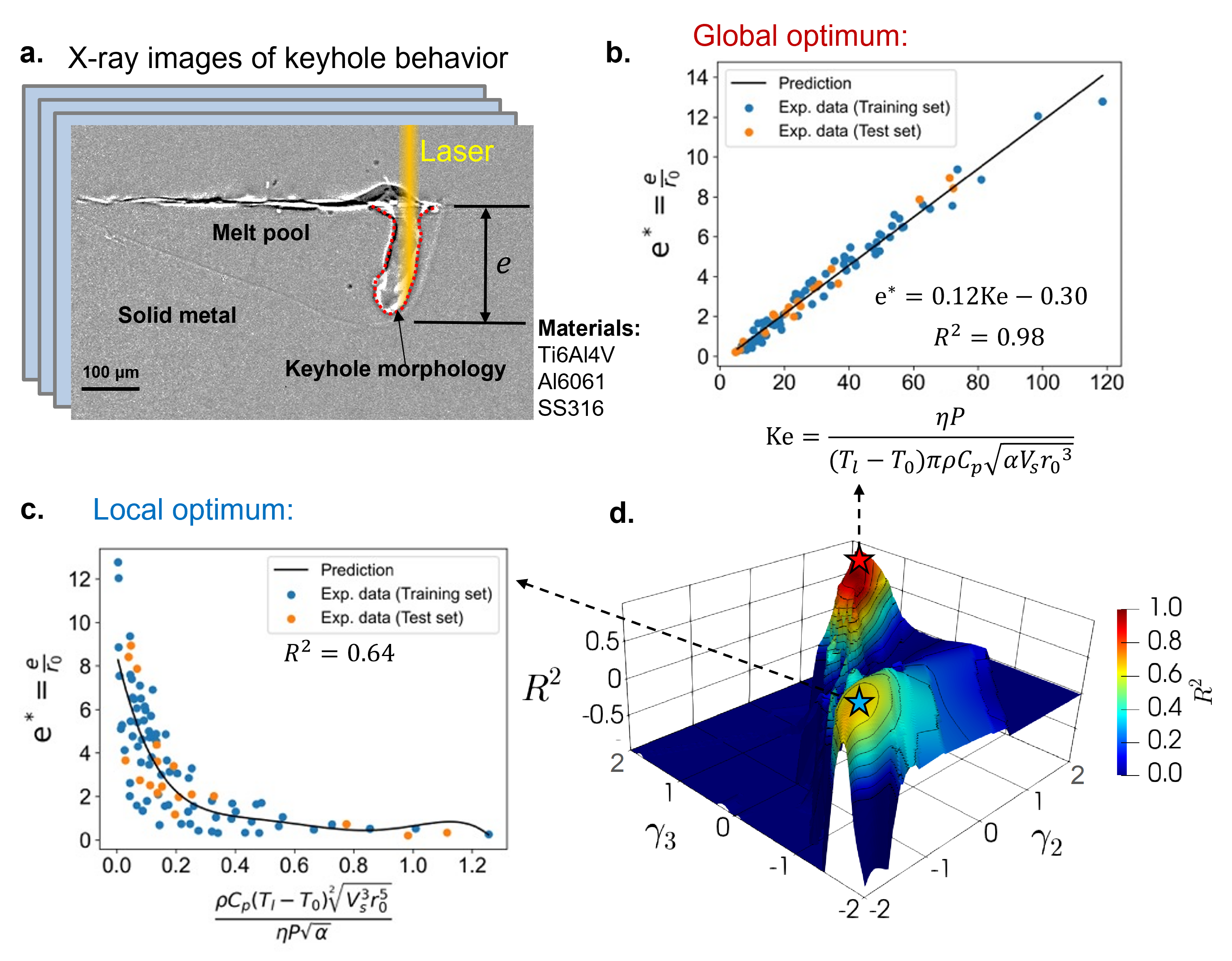}
  \caption{\textbf{Discover dimensionless numbers governing keyhole dynamics in laser-metal interaction.} \textbf{(a)} A illustrative X-ray image of keyhole morphology \cite{gan2021universal}. The dataset including X-ray imaging experiments on three different materials. \textbf{(b)} Global optimum of the dimensionless space, which represents the scaling law between the keyhole aspect ratio and the identified keyhole number using dimensionless learning.  \textbf{(c)} Local optimum of the dimensionless space. \textbf{(d)} Dimensionless space using $\gamma_2$ and $\gamma_3$ as coordinates. The values of $R^2$ indicate the learning performance for the corresponding dimensionless number in the dimensionless space. The values of $R^2$ less than -1 are shown as -1.}
  \label{fig:keyhole_results}
\end{figure}

Recently, high-quality in situ experimental data on keyhole dynamics became available via high-speed X-ray imaging \cite{zhao2017real}. Using X-ray pulses, images of the keyhole region inside the metals can be recorded with micrometer spatial resolution \cite{zhao2020critical}. The keyhole depth $e$ can be measured from the X-ray images (Fig. \ref{fig:keyhole_results}a), and it depends on different materials and several process parameters such as the effective laser power $\eta P$,  the laser scan speed $V_s$, and laser beam radius $r_0$. We collect a dataset of keyhole X-ray images from literatures, including ninety-one experiments with various process parameters and three different materials: a titanium alloy (Ti6Al4V), an aluminium alloy (Al6061), and a stainless steel (SS316) \cite{gan2021universal,zhao2019bulk}. We represent a material using a set of material properties: the thermal diffusivity $\alpha$, the material density $\rho$, the heat capacity $C_p$, the difference between melting temperature and ambient temperature $T_l-T_0$. Therefore, the casual relationship can be expressed as  

\begin{equation}
	e=f(\eta P, V_s, r_0, \alpha, \rho, C_p, T_l-T_0)
	\label{eqn:keyholelist}
\end{equation}

We can use the dimensionless learning described in the previous section to extract a low-dimensional scale-free relation from the parameter list. The dimension matrix $\boldsymbol{D}$ and computed basis vectors $\boldsymbol{w_{b1}}$, $\boldsymbol{w_{b2}}$, and $\boldsymbol{w_{b3}}$ of this example are provided in Supplementary Information Section 1.6. We firstly demonstrate the grid search ranging from -2 to 2 with a interval of 1/100 for the first-level optimization, and 5-ordered polynomial regression for the second-level optimization. We fix $\gamma_1$ as 0.5 and normalize the output variable as the keyhole aspect ratio $e^*=\frac{e}{r_0}$, which is a widely used dimensionless parameter to represent the keyhole characteristic \cite{fabbro2018analysis}. By searching the dimensionless parametric space, we can find one local optimum in term of the $R^2$ criteria marked as a blue star ($R^2 = 0.64$) in Fig. \ref{fig:keyhole_results}d. The expression of the dimensionless number $\Pi = \frac{\rho C_p(T_l-T_0)V_s^{1.5} r_0^{2.5}}{\sqrt{\alpha}\eta P}$ is computed based on the basis coefficients $\gamma_2 = \gamma_3 = -1$. However, the datapoints are scattered shown in Fig. \ref{fig:keyhole_results}c, indicating the dimensionless number located at the local maximum of the dimensionless space is not a good scaling parameter for this problem. The global optimum of the dimensionless space, where $\gamma_2 = \gamma_3 = 1$, provides a much better scaling behavior with a 0.98 $R^2$ score (Fig. \ref{fig:keyhole_results}b). The dominant dimensionless number emerged in the keyhole dynamics is 

\begin{equation}
	\Pi=\frac{\eta P}{(T_l-T_0) \pi \rho C_p \sqrt{\alpha V_s r_0^3}}
	\label{eqn:ke}
\end{equation}

This dimensionless number is identified directly from data, and it has the same form as the newly discovered keyhole number Ke \cite{gan2021universal} (or sometimes called normalized enthalpy \cite{ye2019energy}), which can be derived from heat transfer theory. Even if we use dimensional variable $e$ as the output, the dimensionless learning algorithm still confirms that the form of the keyhole number (i.e., Eqn. \ref{eqn:ke}) is the unique and dominant for controlling the value of keyhole aspect ratio. Details of the procedure and results are provided in Supplementary Information Section 2.1. Using the identified dimensionless number, a simple scaling law emerges to control the keyhole aspect ratio, which simplifies the original high-dimensional problem into an univariate scaling law as
\begin{equation}
	e^*=0.12\mathrm{Ke}-0.30
	\label{eqn:kescalinglaw}
\end{equation}

Providing a sufficient parameter list is critical for the dimensionless learning. If one or more important quantities are omitted, it is impossible to achieve a high $R^2$ for the learning and identify the right form of the dimensionless number(s). We demonstrate in Supplementary Information Section 3.1 that if we assume a parameter list omitting the thermal diffusivity $\alpha$, the maximum $R^2$ is below 0.80 over the dimensionless space, which is much less than the value for the sufficient parameter list (i.e., Eqn. \ref{eqn:keyholelist}). Another scenario often happened in the applications is that we consider more quantities than sufficient, including some irrelevant or unimportant quantities. We demonstrate this scenario in Supplementary Information Section 3.2 by considering one more quantity, e.g., the latent heat of melting $L_m$ or the difference between boiling temperature and ambient temperature $T_l-T_0$, in the parameter list. The form of the keyhole number still can be identified in the scenario. Moreover, there are a few more dimensionless numbers, which consist of the added quantity, providing as high $R^2$ as the keyhole number. It implies that more experiments are needed to select the distinguished one from the identified dimensionless numbers.

We provide two efficient algorithms, i.e., gradient-based and pattern search-based two-level optimization schemes, in Supplementary Information Section 4 to improve the efficiency of the optimization used in this section. These algorithms are especially helpful to explore a high-dimensional parametric space including a lot of parameters to describe the physical system and several dimensionless numbers to construct the low-dimensional pattern.

    
\subsection{Porosity formation in 3D printing of metals.}
\label{porosity_sec}

\begin{figure}
  \centering
  \includegraphics[width=1\linewidth]{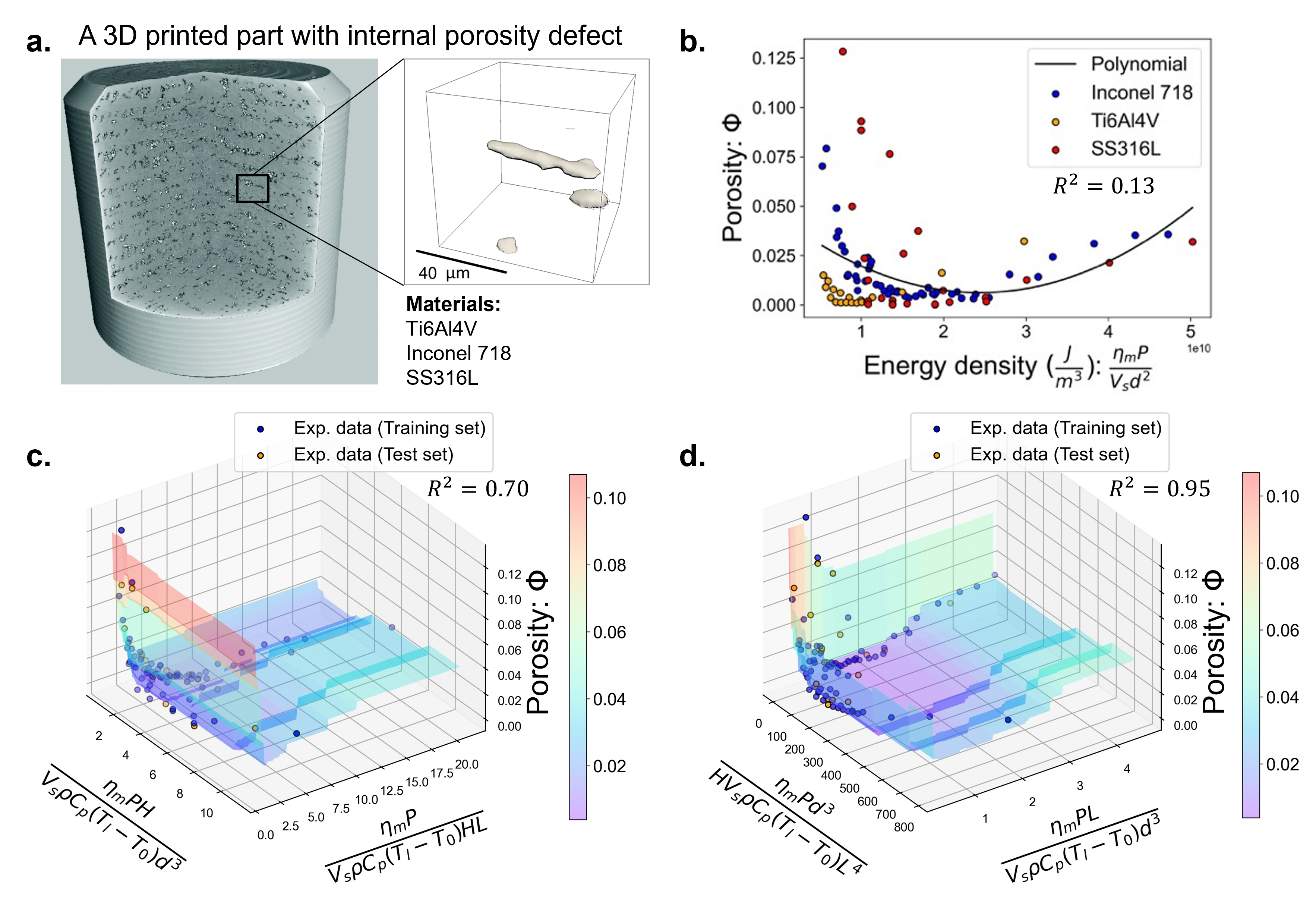}
  \caption{\textbf{Discover dimensionless numbers governing porosity formation in 3D printing.} \textbf{(a)} A schematic of a 3D printed metal part with internal porosity defect \cite{du2018x}. The dataset including X-ray micro-computed tomography (micro-CT) measurements on three different materials. \textbf{(b)} Porosity measurements with various energy density, a traditional combined parameter to correlate porosity with process parameters.  \textbf{(c)} Identified 2D scaling relation combining both lack of fusion and keyhole porosity with two discovered dimensionless numbers \textbf{(d)} Another identified 2D scaling relation with higher $R^2$ score. The reduced parametric space can be easily visualized and interpreted, while the original high-dimensional problem is hard since the porosity is govern by nine parameters.}
  \label{fig:porosity}
\end{figure}

Three-dimensional (3D) printing or additive manufacturing is a  disruptive technology of making three-dimensional solid objects from a digital file, which provides a new paradigm to change manufacturing \cite{dawood20153d}. In metal 3D printing, metallic parts are built layer by layer via local melting and (re)solidification of metallic powders by a laser or electron beam. 3D printing enables remarkable freedom for designing local geometrical and compositional features. However, this process has large numbers of parameters to be considered when making a part, and trends to produce defects such as internal porosity during the process if inappropriate process parameters are used (Fig. \ref{fig:porosity}a).

To extract elegant insights into the complex behavior of the porosity formation in 3D printing, we collect an experimental dataset collected from six independent studies \cite{wang2019dimensionless,kasperovich2016correlation,kumar2019influence,cherry2015investigation,leicht2020effect, simmons2020influence}, including ninety-three 3D printed parts with measured porosity volume fraction and various process parameters. Three kinds of materials were used: a titanium alloy (Ti6Al4V), a nickel-based alloy (Inconel 718), and a stainless steel (SS316L). The porosity volume fraction $\Phi$ depends on many process parameters and materials used in the experiments, which can be expressed as

\begin{equation}
	\Phi=f(\eta_m P, V_s, d, \rho, C_p, \alpha,  T_l-T_0, H, L)
	\label{eqn:porositylist}
\end{equation}

where $\eta_m P$ is the effective laser power input, $V_s$ is the laser scan speed, $d$ is the laser beam diameter, $\rho$ is the material density, $C_p$ is the material heat capacity, $\alpha$ is the thermal diffusivity, $T_l-T_0$ is the difference between melting temperature and ambient temperature, $H$ is the hatch spacing between the two adjacent laser scans, $L$ is the layer thickness of the metallic powders. It is a high-dimensional relation, and thus hard to understand and visualize. Traditionally, some combined parameters, such as energy density $\frac{\eta_m P}{V_s d^2}$, are used to simplify this relation. However, the $R^2$ score of a polynomial model using energy density as input is very low (0.13) as shown in Fig. \ref{fig:porosity}b, indicating that an universal physical relation, which is valid for different materials and processing conditions, cannot be built by using the energy density alone because it is not a scale-free parameter. The form of the relation has to be modified when the energy scale is changed in the experiments with varying process parameters or materials.

We apply the dimensionless learning to this challenging engineering problem and discover some dominant dimensionless numbers that provide an universal physical relation remained accurate for all the experimental conditions. The dimension matrix and computed basis vectors of this example are provided in Supplementary Information Section 1.6. The two-level optimization applied in this example includes a pattern search for the first-level and a XGBoost method to capture the second-level relationships (Supplementary Information Section 2.2). We find that two dimensionless numbers are necessary to represent the dataset since there is no high value of the $R^2$ score (e.g., greater than 0.5) can be achieve if we set only one dimensionless number in the training. A systematic algorithm that determines the number of dimensionless numbers required to govern a physical system is provided in Supplementary Information Section 1.3.

We identify several low-dimensional patterns with the propriety of the scale-free from data. They can achieve high $R^2$ for both the training and test sets (a table summarizing the identified dimensionless numbers are provided in Supplementary Information Section 2.2). Interestingly, we identify another dimensionless number (besides the keyhole number), which has been discovered by theory-driven approach \cite{wang2019dimensionless,gan2021universal}: the normalized energy density NED (Fig. \ref{fig:porosity}c). It can be expressed as 

\begin{equation}
	\mathrm{NED}=\frac{\eta_m P}{V_s\rho C_p(T_l-T_0)HL}
	\label{eqn:ned}
\end{equation}

The normalized energy density NED represents the ratio of laser energy input within the powder layer to sensible heat of melting. This dimensionless number is govern the lack of fusion porosity in metal 3D printing, which is a well-know porosity mechanism due to insufficient laser energy input to fully melt the powder material \cite{du2019effects}. The other dimensionless number $\frac{\eta_m PH}{V_s\rho C_p(T_l-T_0)d^3}$ in Fig. \ref{fig:porosity}c is related to another porosity mechanism, i.e., keyhole porosity, caused by bubbles of gas gotten trapped underneath the surface during the fluctuation of a unstable keyhole \cite{zhao2020critical}. This dimensionless number is a modified normalized enthalpy product, i.e., $\mathrm{NEP}\cdot \frac{H}{d}$, where the normalized enthalpy product NEP is proven to be related to the keyhole instability, and a unstable keyhole with high NEP could lead to keyhole pores \cite{ye2019energy}. The NEP can be expressed as 

\begin{equation}
    \mathrm{NEP}=\frac{\eta_m P}{V_s\rho C_p(T_l-T_0)d^2}
	\label{eqn:nep}
\end{equation}

Since the NEP is derived from the single-track laser scan condition \cite{ye2019energy}, the modified term $\frac{H}{d}$ emerges to consider the effect of multiple-track scanning. Another identified low-dimensional pattern that can achieve even higher $R^2$ (0.95) is shown in Fig. \ref{fig:porosity}d. It is also based on NED and NEP, but with different modification terms. By reducing high-dimensional parameter space, much less experiments would be required to determine optimal processing conditions and parameters for new materials and thus ease the Edisonian burden endemic among current metal 3D printing practitioners.


\section{Discussions}

\begin{figure}
  \centering
  \includegraphics[width=1\linewidth]{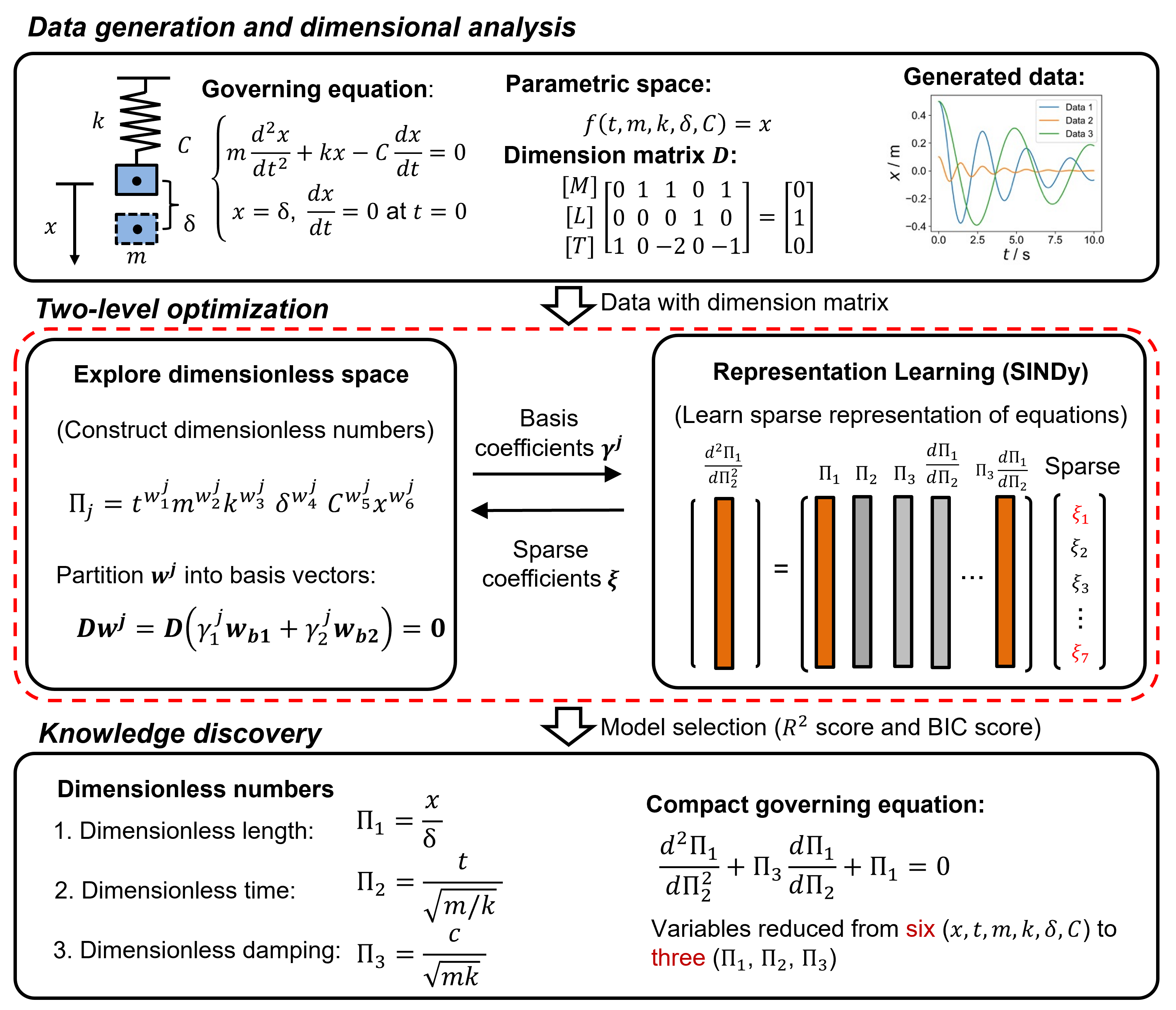}
  \caption{\textbf{Recover dimensionless numbers and the governing equation of a spring-mass-damping system.} In the first step, we generate three sets of data based on the governing equation with different combinations of parameters (spring constant $k$, mass $m$, initial displacement $\delta$, and damping coefficient $c$). After analyzing the parametric space, we can obtain the dimension matrix $\boldsymbol{D}$. In the second step, a two-level optimization is developed to find the best basis coefficients and sparse coefficients sequentially. In third step, we can obtain the best dimensionless numbers and dimensionless governing equation after model selection using $R^2$ score and Bayesian information criterion (BIC) score. Detailed information is shown in Supplementary Information Section 5.}
  \label{fig:pde}
\end{figure}

The two-level optimization scheme makes the dimensionless learning very flexible. The first-level scheme guarantees the dimensional invariance (or called dimensional homogeneity), and thus many representation learning methods can be used for the second-level scheme to capture scale-free relationships. We demonstrate polynomial and tree-based XGBoost \cite{chen2016xgboost} methods in the previous sections. However, the capability of the dimensionless learning can be improved by leveraging more methods, including deep neural networks \cite{hornik1989multilayer}, symbolic regression \cite{schmidt2009distilling}, and Bayesian machine learning \cite{barber2012bayesian}. Another highly promising candidate is the sparse identification of nonlinear dynamics (SINDy) \cite{brunton2016discovering}, which can identify ordinary differential equation (ODE) or partial differential equation (PDE) from data. By integrating the proposed dimensionless learning with SINDy, the forms of dimensionally-homogeneous differential equations can be discovered directly from data, which provides even more insights and interpretations of the physical system. We demonstrate this using a simple example of a spring-mass-damper system in Supplementary Information Section 5. Using the proposed two-level optimization with SINDy to analysis the time series of the spring-mass-damper system, we can obtain not only the governing equation, but several key physical parameters such as the natural time scale (i.e., the inverse of natural frequency), natural length scale and dimensionless damping coefficient (Fig. \ref{fig:pde}). The dimensionless form of the governing equation involves the reduced number of variables from six to three, which is the minimalistic representation of the system. 

One important benefit for embedding dimensional invariance as a physical constraint into machine learning algorithms is that it reduces the learning space by eliminating the strong dependence of the dimension for different variables. The models can represent features as dimensionless numbers and transform the datapoints into a low-dimensional pattern that is insensitive to units and scales. It significantly improve the interpretability of the representation learning since the dimensionless numbers are physically-interpretable, which enable qualitative and quantitative analysis of the systems of interest. It is worthy noting that other kinds of invariances have been embedded into machine learning algorithms and achieved prominent successes, such as translation invariance in convolutional neural network (CNN), time translation invariance in recurrent neural network (RNN), and permutation invariance in graph neural network (GNN). 

The proposed dimensionless learning enables a systematic and automatic learning of scale-free low-dimensional laws from high-dimensional data. It can be applied to a vast number of physical, chemical, and biological systems to discover new or modify existing dimensionless numbers. Furthermore, it can be combined with other data-driven methods such as SINDy to discover dimensionless differential equations and scaling laws. In material science, the identified compact mathematical expressions provide simple transition rules that translate optimal process parameters from one material (or existing materials) to another (or new materials). The dimensionless learning is able to reduce complex, highly multivariate problem spaces into descriptions involving just a few dimensionless parameters with clear physical meanings, which is particularly useful for the engineering problems including many adjustable parameters with various dimensions or units, such as advanced materials processing and manufacturing, microfluidic flow control for precise drug delivery, solar energy systems design, and modern financial market analysis. 


\end{document}